\documentclass[aps,pra,superscriptaddress,12pt,tightenlines,nofootinbib]{revtex4}
\usepackage{amsmath,amsthm,graphicx,amssymb,verbatim,listings}
\usepackage[T1]{fontenc}     
\usepackage{lmodern}         
\usepackage[ pdftex, plainpages = false, pdfpagelabels,
                 pdfpagelayout = useoutlines,
                 bookmarks,
                 bookmarksopen = true,
                 bookmarksnumbered = true,
                 breaklinks = true,
                 linktocpage=all,
                 pagebackref=false,
                 colorlinks = true,
                 linkcolor = BrickRed,
                 urlcolor  = blue,
                 citecolor = BrickRed,
                 anchorcolor = green,
                 hyperindex = true,
                 hyperfigures
                 ]{hyperref}
\usepackage[usenames, dvipsnames]{xcolor}
\usepackage{xifthen} 

\newcommand{\ket}[1]{{\ensuremath{\left| #1 \right\rangle}}}
\newcommand{\bra}[1]{{\ensuremath{\left\langle #1 \right|}}}

\newcommand{\arxiv}[2][]{\ifthenelse{\isempty{#1}}{\href{http://arxiv.org/abs/#2}{{\tt arXiv:\allowbreak{}#2}}} {\href{http://arxiv.org/abs/#2}{{\tt arXiv:\allowbreak{}#2 [#1]}}}}
\newcommand{\pirsa}[1]{\href{http://pirsa.org/#1/}{{\tt PIRSA:\allowbreak{}#1}}}
\newcommand{\booktitle}{\textsl}
\newcommand{\hrefdoi}[2]{\href{https://dx.doi.org/#1}{#2}}

\begin{document}

\title{On Relationalist Reconstructions of Quantum Theory}

\author{Blake C.\ Stacey}
\affiliation{Physics Department, University of Massachusetts Boston}

\date{\today}

\begin{abstract}
  Why I'm not happy with how Relational Quantum Mechanics has
  addressed the reconstruction of quantum theory, and why you
  shouldn't be either.
\end{abstract}

\maketitle

One appealing feature of Carlo Rovelli's proposal for ``Relational
Quantum Mechanics''~\cite{Rovelli:1996} is that it offered a challenge
for those who prefer technical work over slinging sentences: the
reconstruction of quantum theory from information-theoretic
principles. This appeal was witnessed by the means through which I
first learned of RQM, a Wikipedia page written by a fan in
2006~\cite{WP:2006} and since trimmed heavily on the grounds that it
said many things not explicitly stated in the literature already. (As
David Mermin once said, ``Writing on Wikipedia is like writing on
water.'') Of course, the idea of rebuilding quantum theory on a better
foundation long predates Rovelli. In some fashion, it is at least as
old as Birkhoff and von Neumann's explorations. We could also mention
Mackey's challenge and its answer by Gleason, and John Wheeler's
suggested research project for his more promising undergraduates:
``Derive quantum theory from an information theoretic
principle!''~\cite{Fuchs:2014}. But Rovelli's article, along with the
2000 Montreal workshop and Hardy's derivation not long
after~\cite{Hardy:2001, Hardy:2016}, served to mainstream the question
for the current century.

A few days after I posted a first version of these notes online
(\url{https://www.sunclipse.org/?p=3016}), Muci\~no, Okon and Sudarsky
independently filed a lengthy critique of RQM on the
arXiv~\cite{Mucino:2021}. As will become clear over the following
sections, I tend to agree with some of their criticisms. However, the
overall tone here will be closer to that of
Pienaar~\cite{Pienaar:2021, Pienaar:2021b}, who evaluates RQM from as
sympathetic a position as possible, taking as given the legitimacy of
``Copenhagen-ish'' interpretations and seeing how RQM fares when
evaluated in that context. Moreover, my emphasis will be different
from that of all these criticisms, as I will focus upon the
\emph{reconstruction} side of Rovelli's proposal. Has it borne fruit?
When later work claimed to find inspiration in it, how close was the
relationship, and did RQM really hold up its end of the deal?

\section{Rovelli's Reconstruction Postulates}
\label{sec:reconstruction}
Rovelli's preliminary stab at reconstructing quantum theory
from information-theoretic principles has some appealing
features but also a few things going against it, which can be
illustrated by quoting the first two postulates that Rovelli proposes.
\begin{itemize}
\item[\textbf{R1.}] ``There is a maximum amount of relevant
  information that can be extracted from a system.''

\item[\textbf{R2.}] ``It is always possible to acquire new information
  about a system.''
\end{itemize}
These have a little of the feel of Einstein's postulates for special
relativity, in that they \emph{seemingly} run the risk of
contradicting each other~\cite{Mermin:2005}. However, in special
relativity, resolving this dramatic tension required overhauling our
notions of space and time, whereas it is possible to have \textbf{R1}
and \textbf{R2} coexist in a much more mundane way. For example, the
Spekkens toy model is a theory explicitly founded upon local hidden
variables~\cite{Spekkens:2007}, and both \textbf{R1} and \textbf{R2}
hold true in it.

Strangely, for a derivation motivated by a ``relational''
interpretation of quantum mechanics, there is nothing all that
\emph{strongly relational} about the postulates \textbf{R1} and
\textbf{R2}. True, they refer to the information that one system might
hold about another, but any statement in the theories of information
or probability will be ``about relations'' in this sense. \textbf{R1}
and \textbf{R2} neither lean upon nor imply the \emph{relativity of
  physical facts} that RQM is supposed to endorse. In that sense, they
are no more ``relational'' than the Central Limit Theorem is. Laudisa
and Rovelli say that RQM discards the assumption ``that there are
variables that take absolute values, namely values independent from
any other systems''~\cite{Laudisa:2019}. Yet \textbf{R1} and
\textbf{R2}, both singly and in combination, are perfectly compatible
with the existence of absolute physical quantities.

I will not dwell very long on the rest of Rovelli's derivation, as the
other premises that he asserts are self-confessedly provisional. Two
conceptual points do deserve examination, though. First, even granting
all of the mathematical choices that Rovelli is willing to make in
order to sketch how a future derivation might go, like the presumption
of unitary matrices, some postulate is required to ensure that the
derivation does not land in a subtheory of quantum mechanics that
admits an easy classical emulation. For example, one can take the
entirety of Rovelli's postulates and arrive at the theory of von
Neumann measurements upon a single qubit, a theory for which Bell
provided a perfectly satisfactory local-hidden-variable
model~\cite{Bell:1964}. Rovelli's stated premises would also be
satisfied by the Spekkens toy theory of odd prime dimensions, which
provides a local-hidden-variable emulation of qudit stabilizer states
and operations~\cite{Spekkens:2016}. Without something like a further
additional postulate of nonclassical structure, it seems difficult to
avoid these traps.

The other point of note is that Rovelli expresses the hope that his
suggested third postulate could be derived from a consistency
condition between observers. However, this consistency condition is
ill-posed.  The trouble with ``consistency'' in RQM is fundamental,
and one way or another, the difficulty enters with every attempt to
define what the consistency condition is supposed to be. For instance,
Smerlak and Rovelli write, ``It is one of the most remarkable features
of quantum mechanics that indeed it automatically guarantees precisely
the kind of consistency that we see in
nature''~\cite{Smerlak:2007}. They continue, setting up a scenario
involving two observers and a spin system: ``Let us illustrate this
assuming that both $A$ and $B$ measure the spin in the same direction,
say $z$, that is $n = n' = z$''~\cite{Smerlak:2007}. Immediately, we
have trouble:\ RQM is supposed to reject absolute states, absolute
physical properties and everything like that. \emph{Who gets to say,
  then, that these two directions are the same?}

In RQM, the quantum state of a system $S$ with respect to another
system $S'$ is an expression of the relation between $S$ and $S'$. The
subject matter of quantum theory is taken to be ``information'' about
such relations, how that ``information'' is constrained, how it may
change and so forth. Any standard of consistency expressed in these
terms \emph{can only be} a consistency standard between informational
relations, and by the basic edict to reject absolutes, the
informational relations to which it applies \emph{must} be tied to a
specific observer.

As Pienaar has recently phrased it, RQM suffers from \emph{loose-frame
  loopholes}~\cite{Pienaar:2021}. Every physical statement within RQM
must be relative to some system, but it is often not clear which
system that is meant to be, and a condition that is supposed to apply
relative to one might actually only be meaningful for another. Until
these loopholes are closed, there is little hope that any condition
upon intersubjective agreement could provide a helpful supplement to
\textbf{R1} and \textbf{R2}. Indeed, the conjunction of \textbf{R1}
and \textbf{R2} is by itself so mundane that the further premises
should say that the world is \emph{interesting,} not that everything
works out without surprises.

\section{Convexity and the Meaning of Probability}
\label{sec:convexity}
In this section and the next three following, we will investigate the
assumptions employed in the reconstruction work of H\"ohn and
Wever~\cite{Hoehn:2017a, Hoehn:2017b}. This work credits RQM as a
source of conceptual inspiration, and Rovelli has advertised it as
``particulary successful''~\cite{Rovelli:2021}; a colleague informs me
that Rovelli himself finds it the ``best completed
reconstruction''~\cite{Fuchs:2021}. H\"ohn and Wever present
formalized versions of \textbf{R1} and \textbf{R2}, supplement them
with several further postulates, and derive the quantum theory for
systems comprising collections of qubits. Importantly, H\"ohn
notes~\cite{Hoehn:2017a} that their ``informational approach is
generally compatible with (but does \emph{not} rely on)'' RQM, as well
as the Brukner--Zeilinger~\cite{Brukner:2003} and
QBist~\cite{Fuchs:2014b} interpretations. This raises a significant
question. If we put ourselves in the shoes of someone maximally
sympathetic to RQM, a true devotee of the interpretation, do we
actually have reason to take the mathematical steps of the
reconstruction? Does the inspiration translate into algebraic
specifics? The goal here is not to dispute any of H\"ohn and Wever's
mathematics, but to see how good a foundation RQM itself could
actually provide for their derivations. As argued
elsewhere~\cite{Appleby:2017}, their technical steps are of
independent interest. Rather than delve into the group theory, we will
focus on interrogating axioms and motivations, because philosophy
interprets aggression as strength~\cite{Prescod-Weinstein:2020}. Can a
devotee of RQM honestly claim that H\"ohn and Wever's reconstruction
puts flesh on RQM's bones, or is it more the case that H\"ohn and
Wever have provided the ribs and vertebrae too?

First, we should address a couple ideas in RQM that H\"ohn and Wever
do \emph{not} use. Their reconstruction does not depend upon any
inter-observer consistency condition. Moreover, in their approach, an
``observer'' is a question-asking, information-gathering entity. Thus,
while their reconstruction is a completion of Rovelli's original
suggested program (for collections of qubits), it can only derive the
concept of a quantum state relative to a decision-making agent, not an
arbitrary physical system. This leads us directly into an ambiguity in
the RQM literature. While the primary papers on RQM contain statements
to the effect that a quantum state is a ``bookkeeping'' device for
physical facts, rather than a fact itself, the examples they provide
for ``facts'' are descriptions like ``the spin in the direction $z$
was $\hbar/2$''~\cite{Rovelli:2021}. How much daylight is there
between this and the claim that the ket $\ket{z = +\frac{\hbar}{2}}$
is a fact? In order to consistently hold the position that a quantum
state is a bookkeeping contrivance, the values that a quantum state
can be updated to must also be bookkeeping contrivances. This is a
well-known point regarding epistemic and doxastic interpretations of
quantum theory~\cite{Fuchs:2021b}. We shall touch more below on how
RQM effectively re-ontologizes quantum states, a move that raises
problems that sticking fully with the ``bookkeeping device''
interpretation would avoid~\cite{Pienaar:2021, Brukner:2021}.

What, then, about the premises that the reconstruction does invoke?

Axiomatic reconstructions of the quantum formalism often present the
notion that \emph{the state space should be a convex set} fairly early
in their derivations. The idea is to abstract the fact about quantum
theory that, given two density matrices $\rho$ and $\rho'$ of the same
dimension, then
\begin{equation}
  \rho_\lambda := \lambda \rho + (1-\lambda) \rho'
  \label{eq:mixing}
\end{equation}
is also a valid quantum state, for any $\lambda \in [0,1]$. In these
approaches to quantum theory, one argues for convex combinations being
valid states because an experimenter working on a system can do one
preparation with probability $\lambda$ and the other with probability
$(1-\lambda)$. The variable $\lambda$ is assumed to take values
anywhere in the unit interval. H\"ohn and Wever invoke a postulate of
this type, in the form of choosing one system out of a pair with
probability $\lambda$. (Not all reconstructions give convexity an
explicit bullet point or pull it out as a numbered postulate, but
H\"ohn and Wever do; see Assumption 2 in~\cite{Hoehn:2017a}.)

In order for Eq.~(\ref{eq:mixing}) to make sense as a representation
of the procedure of flipping a coin to choose a preparation, we must
be able to prepare a coin with an arbitrary bias. If we relax this,
then we find ourselves again in the realm of theories that satisfy
\textbf{R1} and \textbf{R2} while being classical. Lacking the full
interval can indeed arise naturally, since \emph{a coin itself must be
  a physical system,} described by the same general theory as all the
others. If the basic objects in our theory have only a discrete set of
possible statistical states, then we can only obtain a discrete set of
mixtures. Say I have a Spekkensian toy bit to which we have ascribed a
maximal information state, and which I plan to measure in some way,
such that my probability distribution over the two possible outcomes
is uniform.  So, I have a ``fair coin'', implemented as an object
described by my overall theory.  Let $S$ be some other system, a
collective of toy bits.  If I get the $+$ outcome on my coin toy bit,
I carry out some preparation procedure on $S$.  If I get the $-$
outcome on my coin, I put $S$ through a different preparation process.
My previsions for $S$ can access the interior of the state-space
geometry, but not continuously, or even densely:\ I'm just picking up
the 50--50 mixtures.  The more coins I use, the more internal points I
can pick up, but with a finite number of toy bits available to use as
coins, I can only get so many of them!

We could close off this path and avoid the Spekkensian trap if we were
sure we had the whole unit interval to work with.  Unfortunately, the
RQM literature provides little guidance regarding the interpretation
of probability theory, particularly when it comes to probabilities
that are not $1/n$ for some $n$. (In contrast, the primary literature
on QBism covers the topic at length --- perhaps exhaustively, if not
persuasively~\cite{Fuchs:2014}.) Laudisa and Rovelli are insistent
that RQM defines information ``in the sense of Shannon'', which they
take to be a ``definition of information that has no mentalistic,
semantic, or cognitive aspects''~\cite{Laudisa:2019}. Such a claim can
only be as good as the presumption that the $p_i$ in $-\sum_i p_i \log
p_i$ have no such aspects. That is, it presumes an interpretation of
probability that grants at least some probabilities an objective
status: relative frequencies, propensities, degrees of logical
implication, et cetera~\cite{Pienaar:2021b}. The RQM literature leaves
the question of which such interpretation to adopt mostly up in the
air. We can make a tentative deduction from the fact that for Laudisa
and Rovelli, ``relative information'' measures ``the difference
between the possible number of states of the combined system and the
product of the number of states of the two systems''. Likewise,
Rovelli has more recently written that ``information'' in \textbf{R1}
and \textbf{R2} means ``nothing else than `number of possible distinct
alternatives'\,''~\cite{Rovelli:2021}. In other words, RQM declares
objectivity at the cost of having all fundamental probability
distributions be flat, so that Shannon's formula reduces to Hartley
and Boltzmann's.

Consequently, it is not clear that RQM gives one free license to
invoke arbitrary values of $\lambda \in [0,1]$. In order to make the
reconstruction go, we have to add specifics to the inspiration, and
whether RQM can actually sustain the technical statement we need is
far from established. Moreover, expunging the ``mentalistic, semantic,
or cognitive aspects'' from probability does the same for quantum
states. If $S$ is a qubit and there is a physically correct
expectation value for $\sigma_x$, for $\sigma_y$ and for $\sigma_z$,
then there is a physically correct quantum state for $S$, not just a
mental bookkeeping device. Once the state is ``determined entirely by
a specific history of interactions''~\cite{Laudisa:2019}, then we can
speak of the state for $S$ relative to $O$ whether or not $O$ is
clever enough to calculate it. The ``propensity'' or ``objective
chance'' inheres in the history of events. (I know I'm not the only
one who has read the RQM papers as saying this!)

Schmid has summarized further reasons why the case for convexity is
even more debatable than suggested above~\cite{Schmid:2019}. Consider
again the scenario of flipping a coin to decide between systems, or
preparations of a system. What rules out the possibility of a causal
influence from the coin? It could be that the procedure $P$ and the
procedure ``flip a coin, obtain heads and perform $P$'' are quite
different physically, perhaps because of physical consequences left by
the coin flip. The set of all lists of lab procedures has no intrinsic
convex structure, and trying to impose such a structure upon it is a
tricky business. Schmid points out that an alternative way to justify
a convexity condition is to regard mixing \emph{inferentially}. For
example, instead of using a coin to decide between two sets of
laboratory procedures, one could consider states of \emph{belief}
about whether the preparation was $P$ or $P'$. The set of these states
does have a natural convex structure, inherited from probability
theory~\cite{Schmid:2020}. Yet now we must confront the question of
whether RQM allows this move. Does RQM's physicalist interpretation of
probability allow for $\lambda$ to be read this way? QBism, for
instance, would run with it (and declare that those ``preparations''
are also probabilities~\cite{DeBrota:2021}). But it's hard to imagine
an RQM devotee wanting to be seen as too QBist.

\section{Continuity}
\label{sec:continuity}
H\"ohn and Wever make two assumptions that are important for making
sure the theory being constructed does not fall into the trap of being
the Spekkens toy model. They invoke a continuity assumption, which
they phrase as,
\begin{quotation}
  \noindent [An observer] $O$'s `catalogue of knowledge' about [a
    system] $S$ evolves continuously and every consistent such
  evolution is physically realizable.
\end{quotation}
In the Spekkens toy model, the pure states form a discrete set. One
can try introducing continuity by enlarging the state space to be the
full convex hull of the pure states, but even then, the states of
maximal knowledge will still be that discrete set. Thus, we rely upon
another assumption, that
\begin{quotation}
  \noindent $O$'s total amount of information about $S$ is preserved
  in between interrogations.
\end{quotation}
Time evolution carries pure states to pure states, in other words, and
such evolutions are continous, meaning that the pure states of the
theory being constructed will form a continuous set as well. In order
to make this assumption precise, H\"ohn and Wever reject the Shannon
definition of ``information'' and derive a new measure, which works
out to be the squared Euclidean length of a generalized Bloch
vector. This derivation is one of the points of independent interest
(the interaction of Shannon and non-Shannon information measures in
quantum theory is known to lead to intriguing
structures~\cite{Stacey:2021}). For present purposes, though, we focus
on the motivations, and whether RQM can drive them.

As H\"ohn and Wever make clear, their assumption of continuity is
necessary to narrow down the choice of information measure to their
desired form.  Given the importance of continuity, then, we must ask
how much support the fundamentals of RQM provide for it. And here
again, we run into trouble. Little is said that is definitive, but
what is said favors discreteness and granularity. For example, Rovelli
writes, ``Facts are \emph{sparse}: they are realized \emph{only} at the
interactions between (any) two physical systems''~\cite{Rovelli:2021}.
And at somewhat greater length~\cite{Rovelli:2018},
\begin{quotation}
  \noindent The question of ``what happens between quantum events'' is
  meaningless in the theory. The happening of the world is a very
  fine-grained but discrete swarming of quantum events, not the
  permanence of entities that have well defined properties at each
  moment of a continuous time.
\end{quotation}
If the only thing that flows in between measurements is
meaninglessness, on what grounds can we make any assumptions about
symmetry groups? At best, continuity is an unmotivated postulate,
conceptually additional to the underlying ideas. Likewise, Rovelli
writes in a popularized treatment of quantum
gravity~\cite{Rovelli:2018b},
\begin{quotation}
  \noindent The ``quantization'' of time implies that almost all
  values of time $t$ \emph{do not exist}. If we could measure the
  duration of an interval with the most precise clock imaginable, we
  should find that the time measured takes only certain discrete,
  special values. It is not possible to think of duration as
  continuous. We must think of it as discontinuous: not as something
  that flows uniformly but as something that in a certain sense jumps,
  kangaroo-like, from one value to another.
\end{quotation}
If quantum gravity is a world of sudden jumps, then quantum theory
must encompass them, and its interpretation cannot presume continuity
at a fundamental level. Where, then, do we get the smoothness we need?

We will return to the question of continuity between measurements
during the Discussion section, but first, we need to explore how RQM
handles measurement events themselves.

\section{All Measurements Allowed? Revenge of the Shifty Split}
\label{sec:shifty}
The penultimate numbered postulate of H\"ohn and Wever is the
\emph{unrestrictedness of questions}: ``Every question which yields
legitimate probabilities for every way of preparing [a system] $S$ is
physically realizable by [an observer] $O$.'' I would argue that this
is a good strategy. Quantum theory is very broadly applicable, and a
sensible way to understand such a general theory is to start with
assumptions that imply the fewest restrictions and then impose
additional constraints only when strictly necessary. If we wish to
economize on postulates, then we should first see what structures
arise out of minimal sets of them, and only later truncate those
structures. But RQM gets in the way of this! To see why, we will have
to delve into how RQM treats measurements, and the timing thereof.

A commonplace criticism of ``the Copenhagen interpretation'' is that
it leaves unspecified when the process of ``measurement'' takes over
from unitary, Schr\"odingerian time evolution. According to this
critique, subscribing to ``the Copenhagen interpretation'' is rather
like saying that most of the time, gravity is an inverse-square law,
but it stochastically switches to an inverse-cube dependence for brief
moments. This would be, of course, a pathological feature for a theory
to have. It is also rather disconnected from what Bohr actually
wrote. To a Bohrian, there is not a sudden shift between different
dynamical laws, but instead a contextual change in what language can
be applied, in particular regarding when a system should be treated in
functional or in structural terms~\cite{Schlosshauer:2015}. Most
likely, a Bohrian would say that in order to communicate the result of
an experiment to another physicist, we would of course have to
describe it unambiguously, and any sufficiently unambiguous
description --- ``In natural language, potentially augmented by the
concepts of classical physics,'' they might say --- would necessarily
fix any dividing line in place. (A Heisenbergian would instead have a
different kind of dividing line in mind, and say that it \emph{can} be
shifted, but without operational
consequence~\cite{Schlosshauer:2015}. Still another species of
Copenhagener, a Peresian, would agree with what the Bohrian said about
languages, and then argue that we can move the quantum-classical
``cut'' under some conditions, never without consequence but sometimes
with effects that are statistically negligible~\cite{Peres:1993}.)
Everettian interpretations of quantum mechanics shunt the vagueness of
the term ``observer'' over to the question of when wavefunction
branches are ``separate''~\cite{Peres:1993}. RQM washes its hands of
trying to define observers by declaring that any physical system can
serve as one, which as we will see just transfers the load to the
definition of \emph{interaction.}

The primary literature on RQM carries forward the ahistorical
idea~\cite{Chevalley:1999} that ``the'' Copenhagen interpretation is
well-defined and identifiable with what is found in
``textbooks''~\cite{Laudisa:2019}. It also opens the question of
whether RQM actually manages to evade the critiques leveled at
whatever the people out to ``destroy the Copenhagen interpretation''
imagine it to be. (Declarations that ``the'' Copenhagen interpretation
must be demolished are as plentiful as they are historically and
conceptually muddled~\cite{Fuchs:2018}.) The \booktitle{Stanford
  Encyclopedia} article by Laudisa and Rovelli raises this very point
and walks into a loose-frame loophole:
\begin{quotation}
  \noindent The history of a quantum particle, for instance, is
  neither a continuous line [in] spacetime (as in classical mechanics),
  nor a continuous wave function on spacetime. Rather, with respect to
  any other system it is a discrete set of interactions, each
  localized in spacetime.

The flash ontology of RQM seems to raise a difficulty: what determines
the \emph{timing} for the events to happen? The problem is the
difficulty of establishing a specific moment when say a measurement
happens. The question is addressed in Rovelli (1998), observing that
quantum mechanics itself does give a (probabilistic) prediction on
when a measurement happens. This is because the meaning of the
question whether or not a measurement has happened is to ascertain
whether of not a pointer variable $O_A$ in the observing system $S$
has become properly correlated with the measured variable $A$ of the
system $A$. In turn, this is a physical question that makes sense
because it can be posed empirically by measuring $A$ and $O_A$ and
checking if they are consistent.
\end{quotation}
Here, we have the puzzling situation that what should be the most
fundamental scenario, one system observing another, can only be said
to make sense if we bring in a third party. Yet why should properties
that exist relative to that third party be at all binding upon the
first two systems? The joint state of the first two relative to the
third is, by definition, not the quantum state of the second relative
to the first. But it is the latter which changes ``when a measurement
happens''. (This is one place where the concerns in my original notes
largely parallel the critique by Muci\~no et
al.~\cite{Mucino:2021}. However, their further statement that ``the
notion of measurement cannot be part of [the quantum] framework at the
fundamental level'' seems a nonstarter to me, rather like demanding
that game theory cannot have ``player'' as a basic concept~\cite[\S
  17]{DeBrota:2018}. The years have taught me that this is a more
fundamental disagreement than I can hope to resolve within a
parenthesis, and so I will simply remind the reader that, like
Pienaar~\cite{Pienaar:2021}, I am trying to evaluate RQM from the most
generous perspective that I possibly can.)

Smerlak and Rovelli's 2007 paper ``Relational
EPR''~\cite{Smerlak:2007} indicates that RQM finds the Heisenberg
picture of time evolution ``far more natural'' than the
Schr\"odinger. In a way, this is an invigorating move:\ It is tempting
to wonder what attitudes physicists would find natural if we were
taught all along that observables always change smoothly and
unitarily, while states only ever change suddenly and
stochastically. However, it does not seem obligatory. Whatever one
does with the expectation values $\bra{\psi} U_t^\dag A U_t
\ket{\psi}$, one must put the unitaries somewhere, and per the ethos
of denying absolutes, none of the ingredients in that expression
should be taken as intrinsic to the observed system.  Overall, the
important thing is that the relative physical condition of the
observer and the observed system must vary with $t$, regardless of
which mathematical entity is chosen to stash that dependence in.

The term \emph{flash ontology} suggests rather strongly that the
``flashes'' are ontological. Something is realized in the flash:\ A
variable takes a physical value, even if only in relation to a single
other system. This is at odds with the treatment in Rovelli's 1998
paper to which the \booktitle{Stanford Encyclopedia} article
refers~\cite{Rovelli:1998}. In that paper, Rovelli declares,
\begin{quotation}
  \noindent I am not claiming that there is an ``element of reality''
  in the fact that a measurement has happened, or that, in general
  ``measurement having happened or not'' is an \emph{objective
    property} of the coupled system.
\end{quotation}
Yet what are the ``flashes'' supposed to be, other than elements of
relational reality? Moreover, the ``time of measurement'' calculated
in the 1998 paper is not the time at which an observer $O$ will have
measured a system $S$ (obtaining, one might guess, a
``flash''). Instead, it is the time at which the joint state of the
combined $SO$ system, relative to a second observer $O'$, will take a
particular form. Using this as the means to define timing would mean
that only an observer who does not experience the flash can have
timing information about it. This is difficult to square with the idea
that the state of system $S$ relative to observer $O$ is supposed to
encapsulate the history of the interactions between them. Laudisa and
Rovelli declare that any quantum state is ``nothing more than a
compendium of information'' that is ``determined entirely by a series
of interactions:\ the interactions between the system and a second
`observing' system''~\cite{Laudisa:2019}. In their view, a quantum
state ``codes the values of the variables of the first that have been
actualised in interacting with the second''~\cite{Laudisa:2019}. If
this is the case, then the quantum state of $S$ relative to $O$ will
change with each flash, except in the measure-zero set of
circumstances where it was already an eigenstate of the observable
whose value became actual in the flash. But RQM provides no way to
predict, model or even define the times at which these changes
occur:\ The only account of measurement timing it can offer is
relative to a second observer $O'$, who is not a party to the flashes
between $O$ and $S$. Thanks to this loose-frame loophole, RQM leaves
the manner and mode of time evolution essentially underdetermined. The
most basic criticism that anti-Copenhagenists aim at what they call
``Copenhagen'' hits RQM dead center.

RQM is ambiguous on whether some ``observables'' are physically
preferred or not. If any basis can correspond to a valid measurement,
then one is entitled to say that a measurement has occurred at
\emph{any} time. On the one hand, the RQM literature insists that
there is no special role for human observers or any particular kind of
measuring apparatus. On the other hand, Rovelli dodges the
preferred-basis problem by invoking just such a
distinction~\cite{Pienaar:2021}. We can do no better than to quote
Rovelli at this point~\cite{Rovelli:1996}:
\begin{quotation}
  \noindent [G]iven an arbitrary state of the coupled $S$--$O$ system,
  there will always be a basis in each of the two Hilbert spaces which
  gives the bi-orthogonal decomposition, and therefore which defines
  an [operator on the Hilbert space of $S$--$O$] for which the coupled
  system is an eigenstate. But this is of null practical nor
  theoretical significance. We are interested in \emph{certain}
  self-adjoint operators only, representing observables that we know
  how to measure; for this same reason, we are only interested in
  correlations between \emph{certain} quantities: the ones we know how
  to measure.
\end{quotation}
In order to avoid having a vacuous interpretation, Rovelli
\emph{denies} a key assumption of the ``best completed
reconstruction'' inspired in part by RQM. Trying to put flesh on the
bone, we get organ rejection instead! Until this is resolved, a
quantum reconstruction by way of RQM principles has no good reason to
presume that all questions are physically permitted.

\section{Locality}
\label{sec:locality}
In the last step of their argument, H\"ohn and Wever invoke
\emph{tomographic locality,} the idea that the joint state of a
bipartite system is fully characterized by statistics for measurements
on the two parts and their correlations. In other words, only
measurements on the individual halves are necessary, rather than
``global'' measurements of the pair together. As several other
reconstruction programs have done, H\"ohn and Wever use this postulate
to rule out real-vector-space quantum theory. Mathematically, it is
successful in that regard.

But as Wootters has asked~\cite{Wootters:2013}, if local tomography
\emph{failed,} would we really see that failure as much stranger than,
say, entanglement? Moreover, if a local account of quantum physics is
possible in RQM, it is likely to resemble that provided by QBism, and
since the QBist account of physical locality does not demand local
tomography, it is doubtful that RQM could either.

QBism is a local interpretation of quantum mechanics, firstly because
it ties sample spaces to the agent~\cite{Fuchs:2016}. That which
changes ``at a distance'' in a Bell scenario, for example, is only the
agent's expectations about what might happen if she were to travel to
that distant location and take action there. Moreover, these changes
of expectation require no \emph{signalling} to account for. For a
QBist, the idea of spacetime is a conceptual tool that an agent can
use in navigating the flux of life; it is part of the character of
quantum theory that no change of state requires the introduction of
ontological elements that carry information faster than light, as
plotted on any agent's spacetime diagram~\cite{Mermin:2018}.

This more physical kind of locality is harder to abandon than the
tomographic definition, and the latter could be discarded while
keeping the former. For this reason (among others), QBist work on
quantum reconstruction has avoided relying upon the assumption of
tomographic locality~\cite{Fuchs:2016b}.

The status of physical locality in RQM is somewhat
obscure~\cite{Pienaar:2019}. However, it does seem that if we want RQM
to be a local interpretation, some account \emph{like} the QBist one
is the best that RQM could hope for. Whether such an account might be
available is up for debate; the relentless physicalization of
probabilities in RQM makes some of the moves available to the QBist
inadmissible. Suppose that the joint state of two systems $S$ and $S'$
relative to an observer $O$ is maximally entangled. Then the
realization of a variable of $S$ immmediately implies a probability-1
prediction about the corresponding variable of $S'$, and so \emph{an
  aspect of the physical relation between $S'$ and $O$ must have
  changed.} Smerlak and Rovelli would have it that this is only a
``subjective'' change, analogous to a reader's information about a
distant country changing when they read a newspaper article about
it~\cite{Smerlak:2007}. But this runs headlong into RQM's insistence
that information is \emph{factive,} always to be understood as
objectively present within a physical relation~\cite{Pienaar:2021b}.

Even if we grant RQM an account of physical locality along the lines
of the QBist one, RQM would lack fundamental grounds to posit local
tomography. Indeed, given the ambiguity of physical locality in RQM,
the interpretation seems if anything friendly to abandoning that
assumption.

Might we exchange local tomography for a more justifiable premise? (As
Wilce has noted, there is a certain ``\`a la carte'' aspect to quantum
reconstructions, in that axioms can sometimes be swapped out for one
another~\cite{Wilce:2018}.) Barnum, M\"uller and Ududec have presented
an axiomatic derivation of the quantum formalism with a single-system
focus~\cite{Barnum:2014}. In this approach, later refined by Barnum
and Hilgert~\cite{Barnum:2019}, local tomography is replaced by the
\emph{observability of energy.} Essentially, Barnum and coauthors ask
that time-evolution maps be continuous and reversible, and that the
generator of any such evolution be an observable, conserved
quantity. Could RQM avail itself of such an assumption? Given the
concerns we explored in the previous two sections, the outlook is
cloudy.

It is also possible to select complex vector spaces over real ones by
imposing the condition that a complete set of equiangular lines exist
in the four-dimensional case, which amounts to assuming that a
particular type of joint measurement is possible upon a pair of
elementary systems~\cite{Stacey:2021}. The RQM literature's overall
tendency to identify ``measurements'' with orthonormal bases rather
than more general POVMs would incline against a condition of this type.

\section{Discussion}

There is a fundamental tension in RQM, as the official writings have
had it. On the one hand, the ontology is built from events, and
properties instantiate in flashes. Being is discontinuous, stochastic
and staccato. On the other hand, probabilities are factive,
physicalized, made into properties that sway with all the smoothness
of Maxwellian waves. To the extent that the RQM literature's
discussion of event timing has any content --- that is, assuming the
loose-frame loopholes can somehow be closed --- it tells us that
probabilities evolve continuously, and so, something of a material
kind must always be rolling smoothly. Yet, elsewhere, the physical
stuff of nature is said to hop, skip and jump. Does nature \emph{facit
  saltum,} or not?

Rovelli has written that loop quantum gravity has the advantage over
string theory that the former ``addressed upfront the problem of
describing the fundamental degrees of freedom of a theory without a
fixed background spacetime'', while the latter treats this in an
``indirect'' way (\cite{Rovelli:2013}, via~\cite{Ritson:2021}). We
should apply the same standard to the reconstruction of quantum theory
itself. The postulates offered by RQM have no purchase upon
Bell--Kochen--Specker phenomena. Only by invoking additional
conditions, whose support from RQM is at best equivocal, can we
eventually arrive at the quantum formalism, from which we can then
prove the violation of a Bell inequality in the standard way. What
does this exercise in indirection teach us? Shouldn't we rather
address \emph{upfront} the idea of a theory without a background of
hidden variables?

My informal impression of the philosophy-of-physics community is that
RQM has generally skated through, drawing more sympathy than
criticism. RQM offers a conceptual tranquilizer, promising that we
will be able to understand quantum mechanics without having to make
ourselves too uncomfortable. However, resting in this tranquil
atmosphere has not provided much in the way of insight. Reconstructing
quantum theory from physical principles could change everything from
how we theorize about black holes to how we write undergraduate
textbooks. Yet the principles in the derivation that Rovelli deems
``particularly successful'' are variously satisfied by classical
theories, unsupported by RQM, or in conflict with it outright. The
tangible benefit of the hard work of reconstruction is the
highlighting of the clouds and contradictions in the RQM literature.

\bigskip

I thank C.\ A.\ Fuchs, P.\ A.\ H\"ohn, N.\ D.\ Mermin, J.\ L.\ Pienaar
and D.\ Schmid for helpful comments on various versions of this paper.

\end{document}